\documentclass[final,5p,times,twocolumn,fleqn]{elsarticle}
\biboptions{comma,sort&compress}
\usepackage{amsmath,amssymb,subcaption,multirow,float,graphicx,ecrc}
\usepackage{ecrc}
\usepackage{xcolor}
\usepackage{ulem}
\volume{00}
\firstpage{1}
\journalname{Journal of Subatomic Particles and Cosmology}
\jid{jspc}
\jnltitlelogo{Journal of Subatomic Particles and Cosmology}

\begin{document}

\begin{frontmatter}

\title{Next-to-Leading-Order Calculation of the Light Tensor $(J^P=2^+)$ Hybrid Correlation Function\tnoteref{t1}}   
\tnotetext[t1]{Prepared for the Proceedings of 29th International Conference in Quantum Chromodynamics (QCD 26),  July 20--24, 2026, Montpellier, France.}

\author[1]{R.T. Kleiv\corref{cor1}}
\address[1]{Department of Physical Sciences (Physics), Thompson Rivers University, Kamloops, BC, V2C 0C8, Canada}
\cortext[cor1]{Speaker and corresponding author.}           
\ead{rkleiv@tru.ca}  

\author[2]{J. Ho}
\address[2]{Department of Physics, Dordt University, Sioux Center, Iowa,
            51250, USA}
            
\author[3]{S. Li}
\address[3]{Institut für Kernphysik, Johannes Gutenberg-Universität Mainz, Mainz, D-55128,Germany}

\author[4,6]{S. Narison}
\address[4]{Laboratoire Univers et Particules de Montpellier, CNRS-IN2P3, Case 070, Place Eugène Bataillon, Montpellier, 34095, France}

\author[5]{T.G. Steele}
\address[5]{Department of Physics and Engineering Physics, University of Saskatchewan, Saskatoon, SK, S7N 5E2, Canada}

\author[6]{D. Rabetiarivony}
\address[6]{Institute of High-Energy Physics (iHEPMAD), Univ. Ankatso, Madagascar}          

\pagestyle{myheadings}
\markright{}

\pagestyle{myheadings}
\markright{ }
\begin{abstract}
We report on the QCD calculations underlying a next-to-leading-order (NLO) QCD Laplace sum-rule analysis of the light tensor $(J^P = 2^+)$ hybrid meson mass and coupling. This article focuses on the calculation of the NLO perturbative and leading-log NLO non-perturbative contributions. The diagrammatic renormalization method is employed, and a renormalization-group approach is used to determine the leading-log NLO corrections to condensate contributions, including the dimension-six gluon condensate, whose renormalization we extend to $n_f$ quark flavors. The NLO contributions provide a systematically improved theoretical description of the light tensor hybrid correlation function, leading to more reliable determinations of its mass and coupling. The phenomenological implications of our results are discussed separately in these proceedings.
\end{abstract}


\end{frontmatter}

\section{Introduction}
\label{sec:intro}
Hybrid mesons contain both valence quarks and gluons (see Refs.~\cite{Dudek:2026hld,Meyer:2015eta,Ketzer:2019wmd,Chen:2022asf,Austregesilo:2023bnn} for reviews). Light quark hybrids are expected to be in the same mass region as conventional mesons and gluonia. Here we focus on the light tensor ($J^P=2^+$) channel, for which there are several experimental candidates whose quark and gluon content is unknown~\cite{ParticleDataGroup:2026aaa}. QCD sum rules (hereafter QSR; introduced in Refs.~\cite{Shifman:1978bx,Shifman:1978by} and reviewed in Refs.~\cite{Reinders:1984sr,Narison:1989aq,Narison:2002woh}) have been used to study tensor gluonia in Refs.~\cite{Novikov:1981xi,Narison:1983ii,Narison:1996fm,LI2024138454}, tensor mesons in Ref.~\cite{Bagan:1988ay}, and mixing between the two in Ref.~\cite{Bagan:1987zc}. Ref.~\cite{Tan:2024grd} studied light hybrids at LO in several channels, including the $J^P=2^
+$ channel. In Ref.~\cite{Ho:2026rqs} we used QSR to determine the mass and coupling of the light tensor hybrid using a different current than the one used in Ref.~\cite{Tan:2024grd}. Our analysis included next-to-leading-order (NLO) perturbative contributions and leading-log NLO non-perturbative contributions. This article focuses on the QCD calculations performed in Ref.~\cite{Ho:2026rqs}. The QSR analysis and its phenomenological implications are discussed in the article by D.~Rabetiarivony in these proceedings.

\section{Current and Correlation Function}
\label{sec:j_pi_qsr}
Light hadronic states with a hybrid component may be probed using a local interpolating operator (current) $J^{\mu\nu}_{qg}$ for which
\begin{gather}
\langle 0 | J^{\mu\nu}_{qg} | 2^+ \rangle \sim \epsilon^{\mu\nu} \, f_{2^+} \,,
\end{gather}
where $f_{2^+}$ (normalized by $f_\pi = 93\,{\rm MeV}$) denotes the coupling of the current to the hadronic state and $\epsilon^{\mu\nu}$ is a polarization tensor. We use the current
\begin{gather}
J^{\mu\nu}_{qg}(x)=g_s \, \bar u(x) \, \sigma^{\mu\alpha}G^{a\,\nu}_\alpha(x) \, T^a d(x) \,,
\label{eq:J}
\end{gather}
where $g_s=\sqrt{4\pi\alpha_s}$ is the strong coupling, $\bar u (x)$ and $d(x)$ are antiquark and quark fields, $\sigma^{\mu\alpha}=\frac{i}{2}\left[\gamma^\mu\,,\gamma^\alpha\right]\,,$ $G^{a\,\nu}_\alpha(x)$ is the gluon field strength tensor, and $T^a$ are $SU(3)_c$ generators. Using Eq.~\eqref{eq:J}, we construct the correlation function
\begin{gather}
\begin{split}
\psi^{\mu\nu\rho\sigma}_{qg}(q) &= i \!\int \! d^4 x \, e^{i q\cdot x} \langle 0 | T\left[\right.\! J^{\mu\nu}_{qg}(x) J^{\rho\sigma\, \dag}_{qg}(0) \!\left.\right] | 0 \rangle
\\
&=P^{\mu\nu\rho\sigma} (q) \, \psi_{qg}(q^2)  \,.
\label{eq:pi}
\end{split}
\end{gather}
The projector isolates the component of $\psi^{\mu\nu\rho\sigma}_{qg}$ that couples to spin-2 states and is defined as
\begin{gather*}
\begin{split}
&P^{\mu\nu\rho\sigma}(q) = \tilde\eta^{\mu\rho}\tilde\eta^{\nu\sigma}+\tilde\eta^{\mu\sigma}\tilde\eta^{\nu\rho}-\frac{2}{d-1}\tilde\eta^{\mu\nu}\tilde\eta^{\rho\sigma} \,,
\end{split}
\end{gather*}
\begin{gather}
\begin{split}
&\tilde\eta^{\mu\nu} = q^2g^{\mu\nu}-q^\mu q^\nu
 \,,
\\
&P_{\mu\nu\rho\sigma} P^{\mu\nu\rho\sigma}=2q^8(d+1)(d-2)\,, 
\label{eq:proj}
\end{split}
\end{gather}
where $d$ is the number of spacetime dimensions~\cite{Chetyrkin:1998yr}. We use dimensional regularization with $d=4+2\epsilon$ and the ${\rm \overline{MS}}$ scheme, with renormalization scale $\nu$. We calculate the correlation function in Eq.~\eqref{eq:pi} using FeynCalc~\cite{Mertig:1990an,Shtabovenko:2016sxi,Shtabovenko:2020gxv,Shtabovenko:2023idz}, Tarcer~\cite{Mertig:1998vk}, LiteRed~\cite{Lee:2012cn,Lee:2013mka}, and loop integrals given in Ref.~\cite{Pascual_and_Tarrach}.

\section{The Correlation Function at LO}
\label{sec:pi_LO}
\begin{figure}[H]
    \centering
    \begin{subfigure}[b]{0.22\textwidth}
        \includegraphics[]{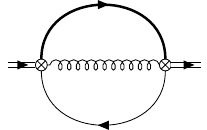}
        \caption{$\psi^{\left(1,1\right)}$}
       \label{fig:diagram11}
    \end{subfigure}
    \begin{subfigure}[b]{0.22\textwidth}
       \includegraphics[]{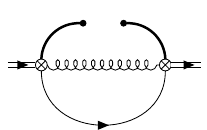}
       \caption{$\psi^{\left(1,2\right)}$}
       \label{fig:diagram12}
    \end{subfigure}
    \\
    \begin{subfigure}[b]{0.22\textwidth}
         \includegraphics[]{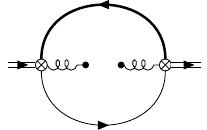}
        \caption{$\psi^{\left(1,3\right)}$}
        \label{fig:diagram13}
    \end{subfigure}
    \begin{subfigure}[b]{0.22\textwidth}
         \includegraphics[]{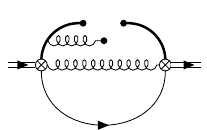}
          \caption{$\psi^{\left(1,4\right)}$}
      \label{fig:diagram14}
    \end{subfigure}
     \\
    \begin{subfigure}[b]{0.22\textwidth}
     \includegraphics[]{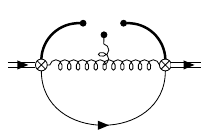}
      \caption{$\psi^{\left(1,5\right)}$}
      \label{fig:diagram15}
    \end{subfigure}
    \begin{subfigure}[b]{0.22\textwidth}
         \includegraphics[]{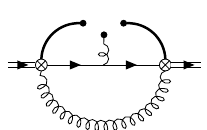}
         \caption{$\psi^{\left(1,6\right)}$}
      \label{fig:diagram16}
    \end{subfigure}
     \\
    \begin{subfigure}[b]{0.22\textwidth}
         \includegraphics[]{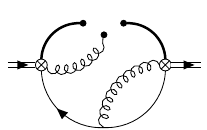}
         \caption{$\psi^{\left(1,7\right)}$}
       \label{fig:diagram17}
    \end{subfigure}
    \begin{subfigure}[b]{0.22\textwidth}
       \includegraphics[]{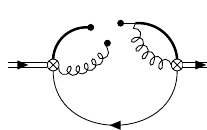}
       \caption{$\psi^{\left(1,8\right)}$}
        \label{fig:diagram18}
    \end{subfigure}
     \\
    \begin{subfigure}[b]{0.22\textwidth}
        \includegraphics[]{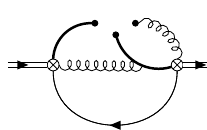}
        \caption{$\psi^{\left(1,9\right)}$}
       \label{fig:diagram19}
    \end{subfigure}
    \begin{subfigure}[b]{0.22\textwidth}
       \includegraphics[]{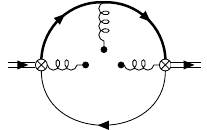}
       \caption{$\psi^{\left(1,10\right)}$}
        \label{fig:diagram110}
    \end{subfigure}
     \\
    \begin{subfigure}[b]{0.22\textwidth}
        \includegraphics[]{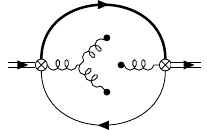}
        \caption{$\psi^{\left(1,11\right)}$}
        \label{fig:diagram111}
    \end{subfigure}
    \begin{subfigure}[b]{0.22\textwidth}
         \includegraphics[]{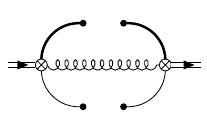}
          \caption{$\psi^{\left(1,12\right)}$}
        \label{fig:diagram112}
    \end{subfigure}
\caption{Feynman diagrams depicting LO OPE terms. Thick and thin fermion lines represent different quark flavors, while $\otimes$ represents an insertion of the light tensor hybrid current. Diagrams with the opposite flavor condensing can be obtained by switching the thick and thin quark lines in these diagrams. Other than flavor exchanges, some diagrams have additional permutations which are not shown here. All Feynman diagrams were drawn using TikZ-Feynman~\cite{Ellis:2016jkw}.}
    \label{fig:LOdiagrams}
\end{figure}

The correlation function in Eq.~\eqref{eq:pi} is calculated in the operator product expansion (OPE), including operators up to dimension-six and Wilson coefficients calculated to LO. Feynman diagrams representing OPE terms are denoted as $\psi^{\left(N,n\right)}$, where $N$ is the order in $\alpha_s$ and $n$ is the diagram number. Each diagram has a multiplicity $M_{(N,n)}$ which is given in Table~\ref{tab:mult}. 
Fig.~\ref{fig:LOdiagrams} shows LO ($N=1$) diagrams.
\begin{table}[H]
\centering 
\renewcommand{\arraystretch}{1.5}
\begin{tabular}{|lll|}
\hline
$N$ & $n$ & $M_{(N,n)}$ \\ 
\hline
$1$ & $7$--$9$ & $2$ \\
$2$ & $2$, $3$, $8$--$12$ & $2$ \\
$2$ & $7$ & $4$ \\
\hline
\end{tabular}
\caption{Multiplicities for LO and NLO diagrams $\psi^{(N,n)}$. If a diagram is not listed its multiplicity is $1$.}
\label{tab:mult}
\end{table}

The LO contribution to the correlation function is
\begin{gather}
\begin{split}
\psi_{qg}^{\rm LO}(Q^2)&=\sum_{n=1}^{12} M_{(1,n)}\psi_{qg}^{(1,n)}(Q^2) 
\\
&=Q^2\Biggl[ \left(d_0 + \frac{d_2}{Q^2} + \frac{d_{4q}}{Q^4} + \frac{d_{6qg}}{Q^6}\right) L  \Biggr. 
\\
&\Biggl.+ 
\frac{d_{4g}+d^c_{4q}}{Q^4}+\frac{d_{6q}+d_{6g}+d^c_{6qg}}{Q^6} \Biggr] \,,
\\
&Q^2 = -q^2  \,, \quad  L=\log\left(\frac{Q^2}{\nu^2}\right)\,.
\end{split}
\label{eq:pi_lo}
\end{gather}

The coefficient functions in Eq.~\eqref{eq:pi_lo} are
\begin{gather}
\begin{split}
&d_0 = \frac{a_s}{864\pi^2}\,, \quad d_2 = \frac{a_s}{80\pi^2}\left(m_u^2+m_d^2\right)\,,
\\
&d_{4q} = \frac{a_s}{24} \left(m_u \langle \bar{u}u \rangle +m_d \langle \bar{d}d \rangle\right) \,, \quad 
d_{4g} = \frac{\langle \alpha_s G^2 \rangle }{72\pi} \,,
\\ 
&d_{6qg} = \frac{7a_s}{432} \left(m_u \langle \bar{u}Gu \rangle +m_d \langle \bar{d}Gd \rangle\right)\,,
\\
& d_{6g} = \frac{\langle g_s^3 G^3 \rangle}{216\pi^2} \,, 
\quad d_{6q} = -\frac{2\pi}{9}\alpha_s \rho  \langle \bar{u}u \rangle  \langle \bar{d}d \rangle \,,
\\
&d^c_{4q} = \frac{a_s}{144}\left[\left(4m_u-3m_d\right)\langle\bar{d}d\rangle + \left(4m_d-3m_u\right)\langle\bar{u}u\rangle\right] \,,
\\
&d^c_{6qg} = \frac{a_s}{32}\left[\frac{125}{81}\left(m_u\langle\bar{u}Gu\rangle+ m_d\langle\bar{d}Gd\rangle\right)\right.
\\
&\qquad\left.-\frac{9}{4}\left(m_u\langle\bar{d}Gd\rangle+ m_d\langle\bar{u}Gu\rangle\right)\right] \,, \quad a_s = \frac{\alpha_s}{\pi}\,.
\label{eq:pi_lo_coeffs}
\end{split}
\end{gather}
The condensates in Eq.~\eqref{eq:pi_lo_coeffs} are defined identically to those in Ref.~\cite{Ho:2026rqs}. 

\section{Laplace sum rules using the LO correlation function}
\label{sec:LSRs_LO}
We construct QCD Laplace sum rules using standard techniques (see e.g. Refs.~\cite{Shifman:1978bx,Shifman:1978by,Narison:1981ts,Narison:2002woh,Bell:1980ub,Bell:1980ww,Bertlmann:1981by}). First, we calculate the second derivative of the correlation function in Eq.~\eqref{eq:pi_lo} with respect to $Q^2$ to construct an Adler-type correlation function that obeys a homogeneous renormalization group equation (RGE). The continuum-subtracted Laplace sum rule moments are given by
\begin{gather}
{\mathcal L}_k(\tau,\nu)
=\int_{t_0}^{t_c}dt~t^k e^{-t\tau}\frac{1}{\pi} \mbox{Im}\,\psi^{\rm had}_{qg}(t,\nu)\,, \quad k\geq 0 \,,
\label{eq:lsr}
\end{gather}
where $t_0$ and $t_c$ are, respectively, the hadronic and continuum thresholds, while $k$ denotes the weight of the moment. The spectral function has been parameterized as a single narrow hadronic resonance and a continuum:
\begin{gather}
\frac{1}{\pi}{\rm Im}\,\psi_{qg}(t)= 2f_{2^+}^2 M_{2^+}^2\delta (t-M_{2^+}^2)+\theta(t-t_c)\rho_n(t_c\,,\tau) \,,
\label{eq:mda}
\end{gather}
where the continuum contribution is modeled as
\begin{gather}
\rho_n(t_c\,,\tau)=e^{-t_c\tau}\sum_{j=1}^n \frac{\left(t_c\tau\right)^j}{j!}\,.
\label{eq:cont}
\end{gather}
We assume that all excited state contributions are smeared by the model above the continuum threshold. Using this resonance model, the ratio of moments
can be used to predict the ground state mass
\begin{gather}
{\mathcal R}_{k+1,k} = \frac{\mathcal{L}_{k+1}}{\mathcal 
{L}_k} \,, \quad  {\mathcal R}_{k+1,k} =  M_{2^+}^2\,.
\end{gather}
Once $M_{2^+}$ has been determined, the lowest weight moment $\mathcal{L}_0$ can be used to predict the coupling $f_{2^+}$.

Using Eqs.~\eqref{eq:pi_lo}~and~\eqref{eq:pi_lo_coeffs}, we find the following sum rule moments:
\begin{gather}
\begin{split}
{\cal L}_{0}^{\rm LO}(\tau)&=\tau^{-2}\Bigg{[}d_0(1-\rho_2) -d_2\tau-(d_{4q}\gamma_E-d_{4q}^c-d_{4g})\tau^2
\\
&+\Big{\{}d_{6g}+d_{6q}+d^c_{6qg}+d_{6qg}(1-\gamma_E)\Big{\}}\tau^3\Bigg{]} \,,
\end{split} \nonumber
\\
\begin{split}
{\cal L}_{1}^{\rm LO}(\tau)&=\tau^{-3}\Bigg{[}2d_0(1-\rho_3) -d_2\tau+d_{4q}\tau^2
\\
&+\Big{\{}\gamma_E d_{6qg}-(d_{6q}+d_{6g}+d^c_{6qg})\Big{\}}\tau^3\Bigg{]} \,,
\end{split} \nonumber
\\
\begin{split}
{\cal L}_{2}^{\rm LO}(\tau)&=\tau^{-4}\Bigg{[}6d_0(1-\rho_4) -2d_2\tau+d_{4q}\tau^2
- d_{6qg}\tau^3\Bigg{]} \,,
\end{split}
\label{eq:srlo}
\end{gather}
 where $\gamma_E$ is the Euler-Mascheroni constant. Notice that in the chiral limit, all coefficient functions vanish, except for $d_0$, $d_{4g}$, $d_{6g}$, and $d_{6qg}$. Also, as the weight of the moment increases, the non-perturbative content of the resulting sum rule tends to decrease. Therefore, it is desirable to calculate higher-order corrections to the perturbative and non-perturbative Wilson coefficients in the OPE.

\section{NLO Perturbative Contributions}
\label{sec:pi_NLO}

\begin{figure}[H]
    \centering
    \begin{subfigure}[b]{0.22\textwidth}
        \includegraphics[]{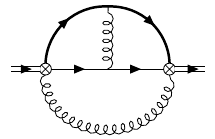}      
        \caption{$\psi^{\left(2,1\right)}$}
       \label{fig:diagram21}
    \end{subfigure}
    \begin{subfigure}[b]{0.22\textwidth}
       \includegraphics[]{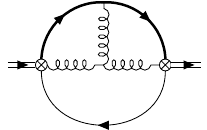}      
       \caption{$\psi^{\left(2,2\right)}$}
       \label{fig:diagram22}
    \end{subfigure}
    \\
    \begin{subfigure}[b]{0.22\textwidth}
        \includegraphics[]{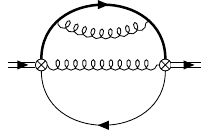}      
        \caption{$\psi^{\left(2,3\right)}$}
        \label{fig:diagram23}
    \end{subfigure}
    \begin{subfigure}[b]{0.22\textwidth}
      \includegraphics[]{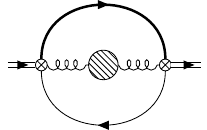}      
      \caption{$\psi^{\left(2,4\right)}$}
      \label{fig:diagram24}
    \end{subfigure}
     \\
    \begin{subfigure}[b]{0.22\textwidth}
    \includegraphics[]{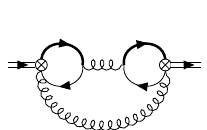}
      \caption{$\psi^{\left(2,5\right)}$}
      \label{fig:diagram25}
    \end{subfigure}
    \begin{subfigure}[b]{0.22\textwidth}
     \includegraphics[]{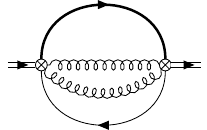}
      \caption{$\psi^{\left(2,6\right)}$}
      \label{fig:diagram26}
    \end{subfigure}
     \\
    \begin{subfigure}[b]{0.22\textwidth}
        \includegraphics[]{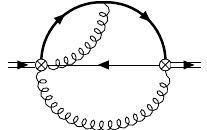}      
         \caption{$\psi^{\left(2,7\right)}$}
       \label{fig:diagram27}
    \end{subfigure}
    \begin{subfigure}[b]{0.22\textwidth}
    \includegraphics[]{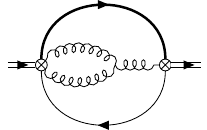}      
       \caption{$\psi^{\left(2,8\right)}$}
        \label{fig:diagram28}
    \end{subfigure}
     \\
    \begin{subfigure}[b]{0.22\textwidth}
    \includegraphics[]{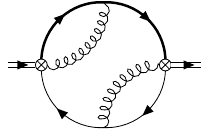}      
       \caption{$\psi^{\left(2,9\right)}$}
       \label{fig:diagram29}
    \end{subfigure}
    \begin{subfigure}[b]{0.22\textwidth}
    \includegraphics[]{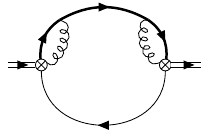}      
    \caption{$\psi^{\left(2,10\right)}$}
        \label{fig:diagram210}
    \end{subfigure}
     \\
    \begin{subfigure}[b]{0.22\textwidth}
        \includegraphics[]{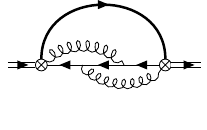}      
        \caption{$\psi^{\left(2,11\right)}$}
        \label{fig:diagram211}
    \end{subfigure}
    \begin{subfigure}[b]{0.22\textwidth}
      \includegraphics[]{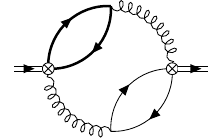}      
      \caption{$\psi^{\left(2,12\right)}$}
        \label{fig:diagram212}
    \end{subfigure}
\caption{NLO diagrams. The shaded loop in diagram $\psi^{(2,4)}$, shown in Fig.~\ref{fig:diagram24}, represents the gluon vacuum polarization, {\it i.e.}, the sum of a quark loop (including $n_f$ chiral flavors), a gluon loop, and a ghost loop. All other notation is identical to that used in Figure~\ref{fig:LOdiagrams}.}
\label{fig:NLOdiagrams}
\end{figure}

NLO perturbative contributions to the OPE are calculated in Feynman gauge and in the chiral limit. The bare NLO diagrams $\psi^{(2,n)}$ shown in Fig.~\ref{fig:NLOdiagrams} can be expressed as
\begin{gather}
    \psi_B^{(2,n)}\left(Q^2\right) = \frac{Q^2 a_s^2}{256\pi^2} \left[ B_0^n \frac{L}{\epsilon} + B_1^n L + \frac{3}{2} B_0^n L^2 \right] \,. 
\label{eq:nlo_pi_b}
\end{gather}
Non-logarithmic terms are omitted in Eq.~\eqref{eq:nlo_pi_b} because they are dispersion relation subtraction constants that vanish upon construction of Laplace sum rules.

To the best of our knowledge, the renormalization of the hybrid current in Eq.~\eqref{eq:J} has not previously been studied. A conventional treatment, such as that of Ref.~\cite{Narison:1983kn}, requires consideration of mixing with all operators of equal or lower dimension. Alternatively, the diagrammatic renormalization method of Ref.~\cite{deOliveira:2022eeq} determines the required counter-terms directly from divergent sub-diagrams. Ref.~\cite{deOliveira:2022eeq} demonstrated the equivalence of the diagrammatic and conventional operator mixing approaches for several different QCD correlation functions, including those for light- and heavy-quark scalar and vector mesons, heavy-light scalar and vector diquarks, and for an off-diagonal scalar quark-gluonium correlation function. The two approaches were also shown to be equivalent for tensor gluonium in Ref.~\cite{LI2024138454}.

\begin{figure}[H]
    \centering
    \begin{subfigure}[c]{0.22\textwidth}
        \includegraphics[]{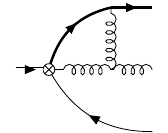}      
        \caption{$\psi^{\left(2,2\right)}_S$}
       \label{fig:diagram22s}
    \end{subfigure}
    \begin{subfigure}[c]{0.22\textwidth}
       \includegraphics[]{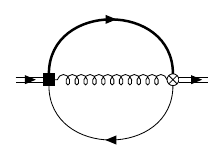}      
       \caption{$\psi^{\left(2,2\right)}_C$}
       \label{fig:diagram22c}
    \end{subfigure}
\caption{Sub-diagram $\psi^{(2,2)}_S$ and counter-term diagram $\psi^{(2,2)}_C$ associated with the left current vertex in Fig.~\ref{fig:diagram22}; those associated with the right current vertex are not shown. The $\blacksquare$ in $\psi^{(2,2)}_C$ represents the divergent part of $\psi^{(2,2)}_S$. All other notation is identical to that used in Figure~\ref{fig:LOdiagrams}.}
\label{fig:psi22sc}
\end{figure}

In the diagrammatic method, the divergent sub-diagrams of each NLO diagram are isolated and used to construct the corresponding counter-term diagrams. We illustrate the procedure using diagram $\psi^{(2,2)}$ in Fig.~\ref{fig:diagram22}, whose bare result is
\begin{gather}
    \psi_B^{(2,2)}\left(Q^2\right) = \frac{Q^2 a_s^2}{256\pi^2} \left[ -\frac{L}{9\epsilon} + \frac{16 L}{15} - \frac{L^2}{6} \right] \,. 
\label{eq:psi22_b}
\end{gather}
Fig.~\ref{fig:psi22sc} shows one of the divergent sub-diagrams of $\psi_B^{(2,2)}$ and its corresponding counter-term diagram.

The $1/\epsilon$ divergent part of sub-diagram $\psi^{\left(2,2\right)}_S$ in Fig.~\ref{fig:diagram22s} can be calculated using the methods outlined in Ref.~\cite{deOliveira:2022eeq}. Accounting for the equivalent sub-diagram associated with the right current vertex, as described in Fig.~\ref{fig:psi22sc}, we find that 
\begin{gather}
    \psi_C^{(2,2)}\left(Q^2\right) = \frac{Q^2 a_s^2}{256\pi^2} \left[ -\frac{L}{9\epsilon} + \frac{107L}{270} - \frac{L^2}{9} \right] \,. 
\label{eq:psi22_c}
\end{gather}
The renormalized diagram $\psi_R^{(2,2)}$ is obtained by subtracting Eq.~\eqref{eq:psi22_c} from Eq.~\eqref{eq:psi22_b},
\begin{gather}
\begin{split}
    \psi_R^{(2,2)}\left(Q^2\right) &= \psi_B^{(2,2)}\left(Q^2\right) -  \psi_C^{(2,2)}\left(Q^2\right)
    \\
    &=\frac{Q^2 a_s^2}{256\pi^2} \left[\frac{181L}{270} - \frac{L^2}{18} \right] \,,
\end{split}
\label{eq:psi22_r}
\end{gather}
The same procedure can be applied to renormalize each individual NLO diagram. All counter-term diagrams can be expressed as
\begin{gather}
\psi_C^{(2,n)}\left(Q^2\right) = \frac{Q^2 a_s^2}{256\pi^2} \left[ C_0^n \frac{L}{\epsilon} + C_1^n L + C_0^n L^2 \right] \,. 
\label{eq:nlo_pi_c}
\end{gather}
The $B$ and $C$ coefficients in  Eqs.~\eqref{eq:nlo_pi_b}~and~\eqref{eq:nlo_pi_c} are given in Table~\ref{tab:BandC}.

\begin{table}[H]
\centering 
\renewcommand{\arraystretch}{1.5}
\begin{tabular}{|rrrrr|}
\hline
$n$ & $B^n_0$ & $B^n_1$ & $C^n_0$ & $C^n_1$ \\ 
\hline
$1$ & $0$ & $-\frac{2}{27}$ & $0$ & $0$ \\
$2$ & $-\frac{1}{9}$ & $\frac{16}{15}$ & $-\frac{1}{9}$ & $\frac{107}{270}$ \\
$3$ & $\frac{8}{81}$ & $-\frac{8}{9}$ & $\frac{8}{81}$ & $-\frac{124}{243}$ \\
$4$ & $-\frac{2}{9}$ & $\frac{59}{27}$ & $-\frac{2}{9}$ & $\frac{31}{27}$ \\
$5$ & $0$ & $0$ & $0$ & $0$ \\
$6$ & $0$ & $0$ & $0$ & $0$ \\
$7$ & $-\frac{2}{45}$ & $\frac{41}{75}$ & $-\frac{2}{45}$ & $\frac{218}{675}$ \\
$8$ & $\frac{1}{3}$ & $-\frac{19}{6}$ & $\frac{1}{3}$ & $-\frac{31}{18}$ \\
$9$ & $0$ & $\frac{4}{1215}$ & $0$ & $0$ \\
$10$ & $\frac{16}{1215}$ & $-\frac{2072}{18225}$ & $\frac{16}{1215}$ & $-\frac{1384}{18225}$ \\
$11$ & $-\frac{2}{135}$ & $\frac{164}{2025}$ & $-\frac{2}{235}$ & $\frac{17}{225}$ \\
$12$ & $0$ & $0$ & $0$ & $0$ \\
\hline
\end{tabular}
\caption{Coefficients $B_0^n$, $B_1^n$ (from Eq.~\eqref{eq:nlo_pi_b}), $C_0^n$, and $C_1^n$ (from Eq.~\eqref{eq:nlo_pi_c}) for diagrams $\psi^{(2,n)}$. Note that $n_f=3$ is used in $\psi^{(2,4)}$ above; expressions for general $n_f$ are given in Ref.~\cite{Ho:2026rqs}. Also note that a factor of 2 is included for $C^n_0$ and $C^n_1$, $n \in \{1, 2, 9, \ldots, 12\}$, to account for the independent (but equivalent) sub-diagrams and their associated counter-term diagrams.}
\label{tab:BandC}
\end{table}
Generalizing the procedure demonstrated in Eq.~\eqref{eq:psi22_r}, the sum of all renormalized NLO perturbative contributions is
\begin{gather}
\psi^{\rm pert}_{qg}\!\left.\right|_{\rm NLO}\left(Q^2\right) = \sum_{n=1}^{12} M_{(2,n)}\left[\psi_B^{(2,n)}-\psi_C^{(2,n)}\right] \,,
\end{gather}
where the diagram multiplicities are given in Table~\ref{tab:mult}. The NLO perturbative terms can be parameterized as 
\begin{gather}
\psi^{\rm pert}_{qg}\!\left.\right|_{\rm NLO}\left(Q^2\right) = 
Q^2\left[ d_0^\prime L + \tilde{d}_0 L^2 \right] \,,
\label{eq:pert_nlo}
\end{gather}
where the coefficient functions are 
\begin{gather}
\begin{split}
&d_0^\prime = \frac{a_s^2}{864\pi^2}\left(\frac{317-720n_f}{1080}\right)
\overset{n_f=3}{=}\frac{ a_s^2}{864\pi^2}\left(-\frac{1843}{1080}\right)  \,,
\\
&\tilde{d}_0 =  \frac{a_s^2}{864\pi^2}\left(\frac{11+6n_f}{72}\right)\overset{n_f=3}{=}
\frac{a_s^2}{864\pi^2}\left(\frac{29}{72}\right)  \,. \label{eq:nlo_pi_renorm}
\end{split}
\end{gather}
The combined LO and NLO perturbative contribution is
\begin{gather}
\begin{split}
&\psi_{qg}^{\rm pert}\left(Q^2\right)= Q^2 \left[ a_s a_{00}L+a_s^2\left(a_{10}L+a_{11} L^2\right)   \right] \,, 
\\
&a_{00}=\frac{d_0}{a_s}\,, \quad a_{10}= \frac{d_0^\prime}{a_s^2} \,, \quad a_{11}= \frac{\tilde{d}_0}{a_s^2}\,, 
\end{split}
\label{eq:a_terms}
\end{gather}
where $d_0$ is given in Eq.~\eqref{eq:pi_lo_coeffs} while $d_0^\prime$ and $\tilde{d}_0$ are given in Eq.~\eqref{eq:nlo_pi_renorm}. Note that for $n_f=3$, $a_{10}/a_{00}\simeq-1.706$ and $a_{11}/a_{00}\simeq0.403$, indicating that NLO contributions are significant but remain under good theoretical control. 

\section{Leading-Log NLO Contributions to Condensates}
\label{sec:NLOcondensates}
As emphasized in Section~\ref{sec:LSRs_LO}, it is desirable to calculate NLO corrections to non-perturbative OPE Wilson coefficients. We focus on $\langle \alpha_s G^2 \rangle$, $\langle g_s^3 G^3\rangle$, and $\alpha_s \langle \bar{u}u \rangle \langle \bar{d} d \rangle$ as their LO Wilson coefficients do not vanish in the chiral limit. Leading-log NLO condensate contributions can be determined using RGE methods, as previously demonstrated for tensor gluonium in Ref.~\cite{LI2024138454}. 
The first step is to extract the anomalous dimension of the correlation function from the NLO perturbative result in Eq.~\eqref{eq:a_terms}. The Adler-type correlation function satisfies a homogeneous RGE: 
\begin{gather}
\begin{split}
&\left[\nu\frac{\partial}{\partial \nu}+\beta\, a_s\frac{\partial }{\partial a_s} -2\gamma^\psi\right]
\frac{d^2\psi^{\rm pert}}{d\left(Q^2\right)^2} =0  \,,
\\
&\beta=\beta_1 \frac{\alpha_s}{\pi}+\ldots\,,
\quad
\beta_1 = \frac{2n_f-33}{6} \,,
\end{split}
\label{eq:adler_RG}
\end{gather}
where $\gamma^\psi$ is the anomalous dimension for the correlation function $\psi_{qg}$. Using the methodology of Ref.~\cite{LI2024138454} and the LO expansions of the anomalous dimension and the $\beta$ function (in the convention of Ref.~\cite{Pascual_and_Tarrach}), we find the LO anomalous dimension of the correlation function to be
\begin{gather}
\begin{split}
\gamma^\psi=\gamma_1^\psi \frac{\alpha_s}{\pi}+\ldots  \,,
\quad
\gamma_1^\psi=
   \frac{\beta_1}{2}-2\frac{a_{11}}{a_{00}} \overset{n_f=3}{=}
   -\frac{55}{18} \,.
\label{eq:RG_beta_gammapsi}
\end{split}
\end{gather} 
The $\langle \alpha_s G^2 \rangle$, $\langle g_s^3 G^3\rangle$, and $\alpha_s \langle \bar{u}u \rangle \langle \bar{d} d \rangle$ OPE terms can be parametrized as
\begin{gather}
\begin{split}
&\psi^{\alpha_s G^2}\left(Q^2\right)=
\frac{1}{Q^2}
\langle \alpha_s G^2 \rangle\left[
b_{00}+ a_s\left( b_{10}+b_{11}L \right)\right] \,,
\\
&\psi^{g^3 G^3}\left(Q^2\right)=\frac{1}{Q^4}\langle g_s^3 G^3\rangle 
\left[
c_{00}+ a_s\left( c_{10}+c_{11}L \right)\right]\,,
\\
&\psi^{\alpha \bar{q}q \bar{q}q}\left(Q^2\right)=\frac{1}{Q^4}
\rho\alpha_s \langle \bar{u}u \rangle \langle \bar{d} d \rangle
\left[
e_{00}+ a_s\left( e_{10}+e_{11}L \right)\right] \,,
\label{eq:b_c_d_terms}
\end{split}
\end{gather}
where the coefficients $b_{00}$, $c_{00}$, and $e_{00}$ can be extracted from Eq.~\eqref{eq:pi_lo_coeffs}. Note that the RGE methodology only determines the leading-log terms, and hence cannot determine $b_{10}$, $c_{10}$, and $e_{10}$. The contribution of each operator in Eq.~\eqref{eq:b_c_d_terms} has the form
\begin{gather}
\begin{split}
&\psi^\chi
 =\langle \chi \rangle \left[ 
 E+a_s\left( F+H\, L \right)
 \right]  \,,  
 \\
 &\langle\chi\rangle \in \left\{ \langle\alpha_s G^2\rangle,\,\langle g^3 G^3\rangle,\, \rho\alpha_s \langle \bar{u}u \rangle \langle \bar{d} d \rangle \right\} \,. 
 \end{split}
 \label{eq:chi}
 \end{gather}
Because the LO coefficient $E$, representing $b_{00}$, $c_{00}$, or $e_{00}$, is finite, Eq.~\eqref{eq:chi} satisfies a homogeneous RGE of the same form as Eq.~\eqref{eq:adler_RG}:
\begin{gather}
\left[\nu \frac{\partial}{\partial \nu} + \beta \, a_s \frac{\partial}{\partial a_s} - 2 \gamma^\psi \right] \psi^\chi = 0 \,.
\label{eq:psichi_rge}
\end{gather}
At LO, the operators in Eq.~\eqref{eq:b_c_d_terms} do not mix under renormalization, so the RGE for the operator $\chi$ is
\begin{gather}
  \nu\frac{d}{d\nu}\langle \chi \rangle=-\gamma_1^\chi \, a_s\, \langle \chi \rangle\,,
 \label{eq:general_RG}
\end{gather}
where $\gamma_1^\chi$ is the LO expression for the anomalous dimension of the operator $\chi$. Substituting Eq.~\eqref{eq:chi} into Eq.~\eqref{eq:psichi_rge}, applying Eq.~\eqref{eq:general_RG}, and neglecting higher order terms, we find that
\begin{gather}
H=-\frac{1}{2}\left(2\gamma_1^\psi+\gamma_1^\chi  \right)E \,.
\label{eq:general_RG_sol}
\end{gather}
Eq.~\eqref{eq:general_RG_sol} determines $b_{11}$, $c_{11}$, and $e_{11}$ once the anomalous dimensions of the corresponding operators are known. 
We therefore consider each operator in turn, beginning with $\langle \alpha_s G^2 \rangle$. Note that mixing between it and $m_q \langle \bar qq\rangle$ occurs at ${\cal O}(a_s^2)$; hence, at LO we have
\begin{gather}
\nu\frac{d}{d\nu }\langle \alpha_s G^2 \rangle=0 \,,
\quad \gamma_1^{\alpha_s G^2}=0\,.
\end{gather}
Using this result along with Eq.~\eqref{eq:general_RG_sol}, we find that
\begin{gather}
b_{11}=-\gamma_1^\psi b_{00} \overset{n_f=3}{=} \frac{55}{1296\pi}\,.
\label{eq:b11}
\end{gather}
Adopting the vacuum saturation approximation, the anomalous dimension for the operator $\alpha_s \langle \bar{u}u \rangle \langle \bar{d}d \rangle$
is related to the $\beta$ function and the quark mass anomalous dimension:
\begin{equation}
 \gamma_1^{\alpha\bar q q \bar q q}  =-\beta_1-2\gamma_1^m\,,
 \quad \gamma^m=-\frac{\nu}{m}\frac{dm}{d\nu} \,, \quad \gamma_1^m=2 \,.
\end{equation}
Substituting these results into Eq.~\eqref{eq:general_RG_sol} yields
\begin{gather}
\begin{split}
&e_{11}=\left(-\gamma_1^\psi+\frac{\beta_1}{2}+\gamma_1^m \right)e_{00} = -\frac{(83+6n_f)\pi}{162}\,,
\\
&e_{11}\overset{n_f=3}{=}-\frac{101\pi}{162}\,.
\label{eq:e11}
\end{split}
\end{gather}
To determine the anomalous dimension of the operator $\langle g_s^3 G^3\rangle$, we consider the related operator ${\cal O}_1$ defined as
\begin{gather}
\begin{split}
\langle g_s^3 G^3\rangle = 4g_s^2 \, \mathcal{O}_1  \,,
\quad
\mathcal{O}_1 =
\frac{g_s}{4} \, \langle f_{abc} \, G_a^{\mu\nu} G_{b\,, \nu\rho} G^\rho_{c\,,\mu} \rangle \,.
\end{split}
\end{gather}
Ref.~\cite{Narison:1983kn} determined the renormalization of $\mathcal{O}_1$ for $n_f=1$, which we generalize here to arbitrary $n_f$. 
The relevant one-loop diagram in Fig.~4 of Ref.~\cite{Narison:1983kn} is flavor independent, allowing its contribution to be extracted from the $n_f=1$ final result. Repeating the renormalization procedure of Ref.~\cite{Narison:1983kn}, with flavor dependence entering through the gluon-field renormalization and hence the $\beta$ function in the background field method, gives
\begin{gather}
\left[ {\cal O}_1\right]_R=Z_{11} \left[ {\cal O}_1\right]_B+Z_{17} \left[ {\cal O}_7\right]_B\,,
\\
Z_{11}=1-\left(
\frac{7}{4}+\frac{n_f}{6}\right)\frac{a_s}{\epsilon}
\,,\quad
Z_{17}=-\frac{9}{8}\frac{a_s}{\epsilon}\,,
\label{Z11}
\end{gather}
where ${\cal O}_7$ is an equation of motion operator whose vacuum expectation value vanishes. Extending the definition of Ref.~\cite{Narison:1983kn} to a sum over $n_f$ quark flavors, the explicit form of ${\cal O}_7$ is 
\begin{gather}
\begin{split}
{\cal O}_7 =& -\frac{1}{2}\left(\partial^\mu G_{\mu\nu,a}+g_s f_{abc} A^\mu_b G_{\mu\nu,c}\right)
\\
&\times\left(\partial^\rho G^\nu_{\rho,a}+g_s f_{amn} A^\rho_m G^\nu_{\rho,n}\right)
\\
&-\frac{1}{2}g_s \left(\partial^\mu G_{\mu\nu,a}+g_s f_{abc} A^\mu_b G_{\mu\nu,c}\right)
\\
&\times\sum_f \bar{\psi}_f\gamma^\nu T^a \psi_f \,.
\end{split}
\end{gather}
 Note that $Z_{17}$ is independent of $n_f$. This follows because the extension of ${\cal O}_7$ to multiple flavors is a linear sum of quark flavors. In the massless limit, the contribution of each flavor can therefore be isolated by choosing Green functions with external quark fields of that flavor. The $n_f=1$ limit of Eq.~\eqref{Z11} recovers the single-flavor result of Ref.~\cite{Narison:1983kn}. The anomalous dimension obtained from $Z_{11}$ is therefore
\begin{equation}
 \nu \frac{d}{d\nu} \langle {\cal O}_1 \rangle  
=-a_s\gamma_1^{{\cal O}_1}  \langle {\cal O}_1 \rangle \,, \quad \gamma_1^{{\cal O}_1}=\frac{7}{2}+\frac{n_f}{3} \,,
\end{equation}
and hence 
\begin{equation}
 \nu \frac{d}{d\nu}  \langle g_s^3 G^3\rangle=-a_s\gamma_1^{g^3G^3}\langle g_s^3 G^3\rangle\,,\quad  \gamma_1^{g^3G^3}  =\gamma_1^{{\cal O}_1}-\beta_1\,.
\end{equation}
Thus, using Eq.~\eqref{eq:general_RG_sol}, we find 
\begin{equation}
 c_{11}=  -\left(\gamma_1^\psi+\frac{1}{2}\gamma_1^{g^3G^3}\right) c_{00}
 \overset{n_f=3}{=} -\frac{13}{1944\pi^2}
 \,.
 \label{eq:c11}
\end{equation}

Combining the NLO perturbative results from Eq.~\eqref{eq:pert_nlo} and the leading-log NLO contributions of the condensates obtained above gives
\begin{gather}
\begin{split}
&\psi_{qg}^{\rm NLO}\left(Q^2\right) =Q^2 \left[ \tilde{d}_0 L^2 + d_0^\prime L
+\left(\frac{\tilde{d}_{4g}}{Q^4}
+\frac{\tilde{d}_{6g}+\tilde{d}_{6q}}{Q^6}\right)L\right] \,,
\\
&\tilde{d}_{4g} = a_s\, b_{11} \,\langle \alpha_s G^2 \rangle \,,
\quad
\tilde{d}_{6g} =a_s \, c_{11} \, \langle g^3G^3\rangle \,,
\\
&\tilde{d}_{6q} =a_s \, e_{11} \,
\rho\alpha_s \langle \bar{u}u \rangle \langle \bar{d}d \rangle,  
\label{eqn:pi_nlo}
\end{split}
\end{gather}
where $d_0^\prime$, $\tilde{d}_0$ are given in Eq.~\eqref{eq:nlo_pi_renorm}, and $b_{11}$, $e_{11}$, and $c_{11}$ are given in Eqs.~\eqref{eq:b11}, \eqref{eq:e11}, and~\eqref{eq:c11}, respectively.

\section{Laplace sum rules with NLO contributions}
\label{sec:LSRs_NLO}
Applying the methods of Section~\ref{sec:LSRs_LO} to the NLO results in Eq.~\eqref{eqn:pi_nlo} gives the following corrections to the Laplace sum rules:
\begin{gather}
\begin{split}
{\cal L}_{0}^{\rm NLO}(\tau)&=\tau^{-2}\Bigg{[}\left\{2\tilde{d}_0(1-\gamma_E)+d_0^\prime\right\}(1-\rho_2) 
\\
&-\tilde{d}_{4g}\gamma_E\tau^2+(\tilde{d}_{6g}+\tilde{d}_{6q})(1-\gamma_E)\tau^3\Bigg{]} \,,
\\
{\cal L}_{1}^{\rm NLO}(\tau)&=\tau^{-3}\Bigg{[}\left\{2d_0^\prime +6\tilde{d}_0\left(1-\frac{2}{3}\gamma_E\right)\right\}(1-\rho_3) 
\\
&+\tilde{d}_{4g}\tau^2+\gamma_E(\tilde{d}_{6g}+\tilde{d}_{6q})\tau^3\Bigg{]} \,,
\\
{\cal L}_{2}^{\rm NLO}(\tau)&=\tau^{-4}\Bigg{[}\left\{6d_0^\prime +22\left(1-\frac{6}{11}\gamma_E\right)\tilde{d}_0\right\}(1-\rho_4) 
\\
&+\tilde{d}_{4g}\tau^2-(\tilde{d}_{6g}+\tilde{d}_{6q})\tau^3\Bigg{]} \,.
\end{split}
\label{eq:LSRs_NLO}
\end{gather}

In Ref.~\cite{Ho:2026rqs}, a QSR analysis using the LO sum rules in Eq.~\eqref{eq:srlo} and the ratio $\mathcal{R}^{\rm LO}_{10}$ predicted a light tensor hybrid mass and coupling of approximately $1.2\,{\rm GeV}$ and $4.9\,{\rm MeV}$, respectively. The analysis was subsequently refined by including the NLO sum-rule corrections above together with the topological charge $\Pi\left(0\right)$, where
 \begin{gather}
\Pi\left(Q^2\right) = Q^4 \psi \left(Q^2\right) \,.
\label{eq:top_ch}
 \end{gather}
The latter was determined in Ref.~\cite{Ho:2026rqs} using a QSR analysis that incorporated the NLO corrections. The resulting predictions for the light tensor hybrid mass and coupling were $2.038\pm0.190\,{\rm GeV}$ and $10.47\pm2.87\,{\rm MeV}$, respectively. The substantial shift from the LO predictions demonstrates the importance of the NLO perturbative and leading-log non-perturbative contributions, together with the topological charge. Moreover, the optimal QSR predictions for these quantities occur at lower values of $\tau$ when NLO contributions are included, suggesting improved OPE convergence.

\section{Conclusions}
\label{sec:conclusion}
We have calculated NLO perturbative and leading-log NLO non-perturbative contributions to the light tensor hybrid correlation function. When incorporated into the QSR analysis together with the topological charge, these corrections produce a substantial shift in the predicted mass and coupling relative to the LO result. The diagrammatic approach streamlines the calculation of NLO perturbative corrections, and RGE methods were used to calculate leading-log NLO non-perturbative corrections. We expect these methods to be equally useful in future NLO QSR studies of hybrid mesons with different quantum numbers or quark flavors.

\bibliographystyle{elsarticle-num} 
 \bibliography{refs}
 
\end{document}